\documentclass[aps,prb,twocolumn,showpacs]{revtex4}

\usepackage{graphicx}

\begin{document}
\draft

\title{Spin Hall effect in a system of Dirac fermions in the honeycomb lattice with
intrinsic and Rashba spin-orbit interaction}
\author{A. Dyrda\l$^{1}$, V. K.~Dugaev$^{2,3}$, and J.~Barna\'s$^{1,4}$
}
\address{
$^1$Department of Physics, A.~Mickiewicz University, Umultowska
85, 61-614 Pozna\'n, Poland \\
$^2$Department of Physics, Rzesz\'ow University of Technology,
Powsta\'nc\'ow Warszawy 6, 35-959
Rzesz\'ow, Poland \\
$^3$Department of Physics and CFIF, Instituto Superior T\'ecnico,
Technical University of Lisbon,  Av. Rovisco Pais, 1049-001
Lisbon, Portugal\\
$^4$Institute of Molecular Physics, Polish Academy of Sciences,
Smoluchowskiego 17, 60-179 Pozna\'n, Poland}
\date{\today }

\begin{abstract}
We consider spin Hall effect in a system of massless Dirac
fermions in a graphene lattice. Two types of spin-orbit
interaction, pertinent to the graphene lattice, are taken into
account -- the intrinsic and Rashba terms. Assuming perfect
crystal lattice, we calculate the topological contribution to spin
Hall conductivity. When both interactions are present, their
interplay is shown to lead to some peculiarities in the dependence
of spin Hall conductivity on the Fermi level.

\pacs{73.43.-f,72.25.Hg,73.61.Wp}
\end{abstract}
\maketitle

\section{Introduction}
Long time ago Dyakonov and Perel predicted, that in some systems
transverse spin current and spin accumulation may be induced by
electric current~\cite{dyakonov,dyakonovlett}. This effect, known
as spin Hall effect (SHE),  is currently extensively studied both
experimentally and theoretically~\cite{hirsch, murakami03, sinova,
kato, kimura, brune} (see also review papers~\cite{engel,
dyakonov08}). The effect may appear in semiconductors as well as
in metals, and originates from spin-orbit interaction. Generally,
such an interaction may be of intrinsic or extrinsic origin. The
extrinsic SHE is associated with scattering mechanisms, like skew
scattering and side jump in the presence of impurities. In turn,
the intrinsic mechanism of SHE is a consequence of an unusual
trajectory of the charge carriers in the momentum space, which may
be described by the Berry phase formalism~\cite{berry,sundram}.
This contribution will be referred to as the topological one.

In this paper we consider the topological contribution to spin
Hall conductivity in a system of Dirac fermions with spin-orbit
coupling. Generally, the form of spin-orbit interaction depends on
the symmetry and structure of the system. The Dirac model turned
out to be useful not only in the relativistic field theory, but
also in condensed matter physics to describe some features of
electronic spectrum (at least in a certain energy range). One of
such systems is two-dimensional graphene, and in this paper we
consider Dirac fermions with the spin-orbit interaction taken in
the form appropriate for graphene and including both intrinsic and
Rashba terms. We believe, that the results derived here  will shed
some light on the spin Hall effect in graphene.

Graphene is a two-dimensional honeycomb lattice of carbon atoms,
with two nonequivalent sublattices. The low-energy electron states
near the K and K' points at opposite corners of the Brillouine
zone can be approximated by the conical energy spectrum. As a
result, charge carriers are described by the Dirac equation
~\cite{katsnelson,geim}. The Fermi surface in a neutral graphene
consists of the nonequivalent points K and K', at which the
valence and conduction bands touch each other. However, when the
intrinsic spin-orbit interaction is included, an energy gap opens
at these points. Unfortunately, it is now believed that the
intrinsic spin-orbit coupling in graphene is rather weak so the
gap is also small. Since the graphene layer is usually on a
substrate, one also can expect spin-orbit interaction of Rashba
type \cite{rashba09}. Moreover, the corresponding coupling
parameter can be tuned externally by a gate voltage. Indeed, a
large Rashba spin-orbit interaction has been reported in a recent
experiment ~\cite{varykhalov08}.

Kane and Mele~\cite{kane} have shown that the intrinsic spin-orbit
interaction opens an energy gap at the Dirac points, and also have
predicted a quantized value of the spin Hall conductivity when the
Fermi level is in the gap. The quantized spin Hall conductivity at
the Dirac points was also confirmed by later analytical and
numerical calculations~\cite{sheng,sinitsyn,onari}. On the other
hand, the presence of Rashba spin-orbit interaction reduces the
gap, and when the Rashba interaction is stronger than the
intrinsic one, the gap becomes closed. Since the magnitude of
intrinsic spin-orbit interaction seems to be significantly smaller
than that assumed originally~\cite{min}, it is possible to reach
the limit opposite  to that considered by Kane and Mele, i.e. the
limit where the Rashba coupling dominates while intrinsic
spin-orbit interaction is negligible. To our knowledge, spin Hall
effect in this limit has not been considered analytically so far.
Therefore, in this paper we present analytical and numerical
results obtained within the linear response theory and Green
functions technique, assuming both intrinsic and Rashba spin-orbit
interaction. We focus on the topological contribution to the
effect assuming perfect crystal lattice and ignoring impurities
and defects, which however may influence the magnitude of spin
Hall effect in real systems \cite{sheng,sinitsyn,onari,qiao}.

The paper is organized as follows. In section 2 we describe the
model and present a general formula for the spin Hall
conductivity. The case of pure intrinsic spin-orbit interaction is
presented in section 3, while the case of Rasba interaction is
described and discussed in section 4. The general case, where both
interactions are present, is described in section 5, while summary
and final conclusions are in section 6.

\section{Model and general formula for spin Hall conductivity}

Including both intrinsic and Rashba spin-orbit coupling, the
effective mass Hamiltonian~\cite{kane} of graphene can be written
in the form
\begin{equation}
\label{1} H=H_{0}+H_{SO}+H_{R}.
\end{equation}
The first term, $H_{0}$, describes the low-energy electronic
states around the Dirac points K and K' in the Brillouin zone, and
has the form
\begin{equation}
\label{2}
H_{0}=\left(
                \begin{array}{cc}
                  0 & v(\pm k_{x} - i k_{y}) \\
                  v(\pm k_{x} + i k_{y}) & 0 \\
                \end{array}
              \right),
\end{equation}
where the upper and lower signs correspond to the points K and K',
respectively, and $v$ is a parameter describing the conical energy
spectrum, $v=\hbar v_F$, with $v_F$ denoting the electron Fermi
velocity. The second term in Eq.(1) describes the intrinsic
spin-orbit interaction in graphene,
\begin{equation}
\label{3} H_{SO}=\left(
                \begin{array}{cc}
                  \pm \Delta_{SO}\sigma_{z}& 0 \\
                 0 & \mp \Delta_{SO}\sigma_{z} \\
                \end{array}
              \right) ,
\end{equation}
with $\Delta_{SO}$ being the relevant parameter ($2\Delta_{SO}$ is
the gap created by the intrinsic spin-orbit coupling in the Dirac
points). Finally, the last term in Eq.(1) stands for the Rashba
spin-orbit term,
\begin{equation}
\label{4} H_{R}=\left(
                \begin{array}{cc}
                       0 & \lambda_{R}(\pm \sigma_{y} +i \sigma_{x}) \\
                  \lambda_{R}(\pm \sigma_{y} -i \sigma_{x}) & 0 \\
                \end{array}
              \right),
\end{equation}
where $\lambda_{R}$ is the corresponding coupling parameter. As in
Eq.(2), the upper and lower signs in (3) and (4) correspond to the
two inequivalent points K and K' of the Brillouin zone,
respectively.

To obtain spin Hall conductivity we introduce first the spin
current density operator,
\begin{equation}
\label{5}
\textbf{j}^{s_{j}}=\frac{1}{2}\left[\textbf{v},s_{j}\right]_{+},
\end{equation}
where $[A,B]_+=AB + BA$ denotes the anticommutator of any
operators $A$ and $B$, $v_i=(1/\hbar )(\partial H/\partial k_i)$
is the velocity operator ($i=x,y$) and $s_j=(\hbar /2)\sigma_j$ is
the $j$-th component ($j=x,y,z$) of the spin operator. Taking into
account the exact form of Hamiltonian $H$ one finds
\begin{equation}
\label{3b} v_x =\pm\frac{v}{\hbar}\left(
                \begin{array}{cc}
                 0 & I \\
                 I & 0 \\
                \end{array}
              \right)
\end{equation}
and
\begin{equation}
\label{3a} v_y =i\frac{v}{\hbar}\left(
                \begin{array}{cc}
                 0 & -I \\
                 I & 0 \\
                \end{array}
              \right) ,
\end{equation}
with $I$ being the $2\times 2$ unit matrix.

In the linear response theory, the dc spin Hall conductivity is
given by the formula~\cite{dyrdal},
\begin{equation}
\label{6} \sigma^{s_{z}}_{xy}=\lim_{\omega\to 0}\frac{e \,\hbar
}{2\omega}Tr\int\frac{d\varepsilon}{2\pi}\frac{d^{2}\mathbf{k}}{(2\pi)^{2}}
\left[v_{x},s_{z}\right]_{+}G_{\mathbf{k}}(\varepsilon +
\omega)v_{y}G_{\mathbf{k}}(\varepsilon),
\end{equation}
where $G_{\mathbf{k}}(\varepsilon)$ is the causal Green function
corresponding to the Hamiltonian (1). This formula will be used in
the following to calculate spin Hall conductivity in some specific
cases as well as in a general situation.

\section{The case of $\lambda_{R} = 0$ and $\Delta_{SO}\ne 0$}

We will consider first the special case, when the Rashba coupling
vanishes, while the intrinsic spin-orbit interaction is nonzero,
$\lambda_{R} = 0$ and $\Delta_{SO}\ne 0$. Such a situation has
been  already considered analytically in the clean
limit\cite{kane}, and also studied numerically in the presence of
impurities\cite{sheng}. Our results are consistent with those
obtained in the above cited works.

When $\lambda_{R} = 0$, Eq.(8) for the spin Hall conductivity
takes the following form:
\begin{eqnarray}
\label{7} \sigma^{s_{z}}_{xy} = \,\int \frac{d^{2}{{\bf{k}}}}{(2
\pi)^{2}}\,\frac{d \varepsilon}{2
\pi}\frac{i\,2\,e\,v^{2}\Delta_{SO}}{[\varepsilon -
\sqrt{\Delta_{SO}^{2} + v^{2} k^{2}} + \mu + i \delta\,{\rm
sign}(\varepsilon)]^{2}}\nonumber
\\
\times\frac{1}{[\varepsilon + \sqrt{\Delta_{SO}^{2} + v^{2} k^{2}}
+ \mu + i \delta\, {\rm sign(\varepsilon)}]^{2}},
\end{eqnarray}
where $\mu$ is the chemical potential.

In the zero temperature limit, the spin Hall conductivity will be
presented in the form
\begin{equation}
\label{8}
\sigma^{s_{z}}_{xy}=\sigma^{s_{z},0}_{xy} \mp \delta \sigma^{s_{z}}_{xy},
\end{equation}
where $\sigma^{s_{z},0}_{xy}$ is the contribution from the fully
occupied valence band, and $\delta \sigma^{s_{z}}_{xy}$ is
associated either with the empty part of the valence band (upper
sign) or with occupied part of the conduction band (lower sign).

The contribution $\delta \sigma^{s_{z}}_{xy}$ takes the form
\begin{eqnarray}
\label{9} \delta \sigma^{s_{z}}_{xy} = \pm\frac{e \Delta_{SO}}{4
\pi} \left. \frac{1}{\sqrt{\Delta_{SO}^{2} + v^{2}
k^{2}}}\right|_{0}^{k_{F}} \nonumber \\ = \frac{e }{4 \pi}\left(1
\pm\,\frac{\Delta_{SO}}{\sqrt{\Delta_{SO}^{2} + v^{2}
k^{2}_{F}}}\right) \, ,
\end{eqnarray}
where the upper sign refers to the situation when the Fermi level
is in the valence band, while the lower one when the Fermi level
is in the conduction band. In turn, the contribution from the
fully occupied valence band can be calculated as
\begin{equation}
\label{10}
\sigma^{s_{z},0}_{xy} = -\frac{e\,v^{2} \Delta_{SO}}{4 \pi}\,
\int_{0}^{\infty} \frac{k\, dk}{(\Delta_{SO}^{2} + v^{2} k^{2})^{3/2}} = -\frac{e}{4 \pi}\,.
\end{equation}

\begin{figure}
\includegraphics[width=0.47\textwidth,angle=0]{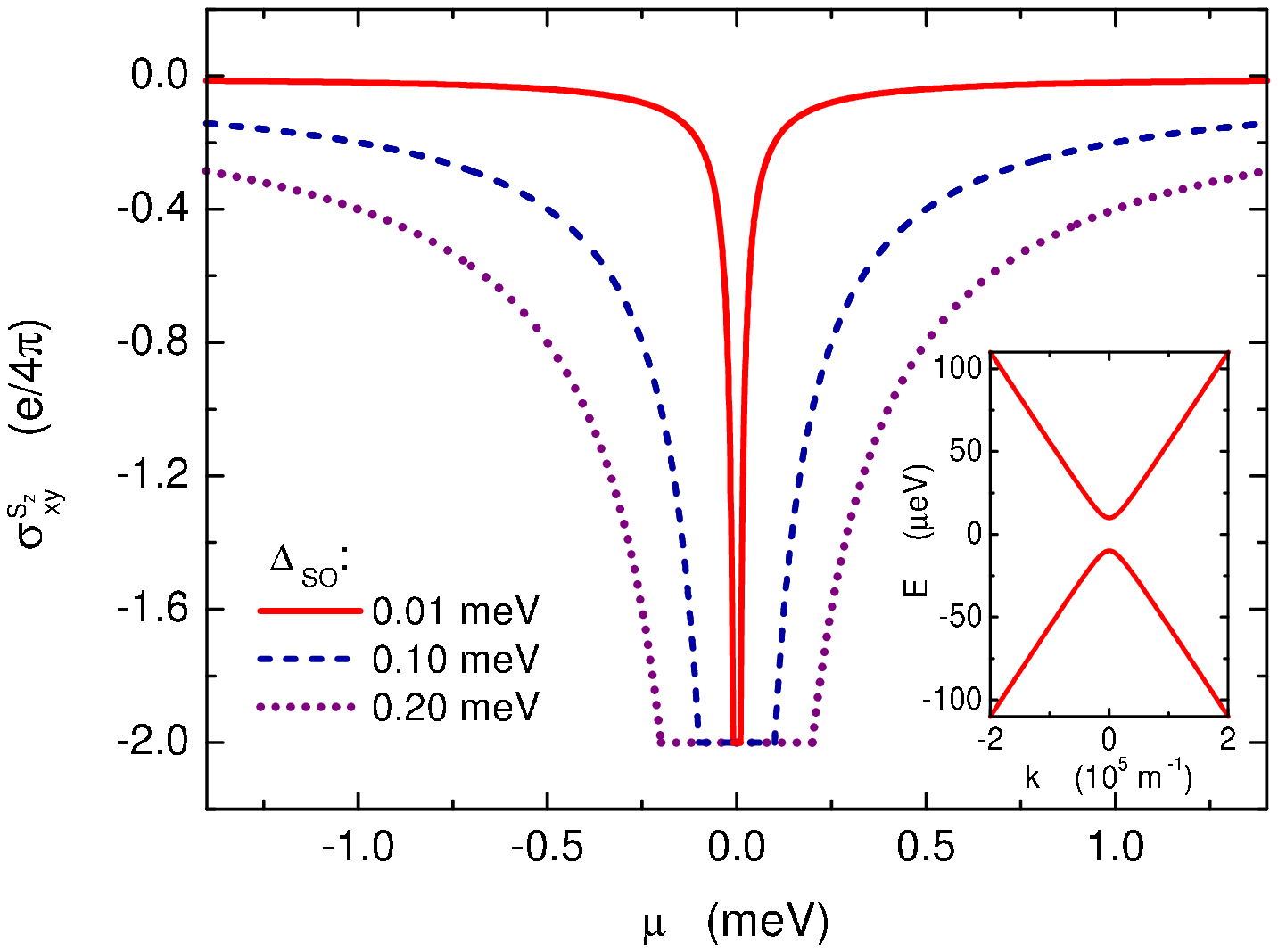}
\caption{(color online) Spin Hall conductivity in the absence of
Rashba interaction and for indicated values of the intrinsic
spin-orbit parameter $\Delta_{SO}$. The inset shows the energy
spectrum in the vicinity of the Dirac point for
$\Delta_{SO}=0.01$meV. The parameter $v$ is defined as
$v=\hbar\,v_{F}$, where $v_{F}$ is the Fermi velocity, $v_{F} =
0.833\times10^{6}$m/s. }
\end{figure}

Thus, from Eqs (\ref{9}) and (\ref{10}) follows that the whole
spin Hall conductivity can be written as
\begin{equation}
\label{11} \sigma^{s_{z}}_{xy} = - \frac{e}{4
\pi}\,\frac{\Delta_{SO}}{|\mu|}
\end{equation}
for $|\mu|>\Delta_{SO}$, and
\begin{equation}
\label{12} \sigma^{s_{z}}_{xy} = - \frac{e}{4 \pi},
\end{equation}
for $|\mu|<\Delta_{SO}$. The above result coincides with the ones
obtained by Kane and Mele\cite{kane} and Sinitsyn et
al~\cite{sinitsyn}. The other two component of the spin Hall
conductivity, i.e. $\sigma^{s_{x}}_{xy}$ and $\sigma^{s_{y}}_{xy}$
vanish, as one could expect.

The corresponding numerical results are shown in Fig.1 for three
different values of the parameter $\Delta_{SO}$. Additional factor
of 2 has been taken into account in order to include the two Dirac
points K and K'. The spin Hall conductivity is shown there as a
function of the chemical potential measured from the middle of the
gap. When the Fermi level is in the energy gap, the Hall
conductivity is constant and quantized. When, in turn, the Fermi
level is either in the valence or conduction bands, the absolute
value of conductivity decreases, and disappears for $|\mu|\to
\infty$. Note, the conductivity is symmetric with respect to the
middle of the gap. The inset in Fig.1 shows the energy spectrum in
the vicinity of the Dirac point, calculated for the parameter
$v_{F}$ obtained by Gmitra et al~\cite{gmitra}.

\section{The case of $\Delta_{SO} = 0$ and $\lambda_{R}\ne 0$}

Now, we consider the opposite situation, i.e. when the  intrinsic
spin-orbit coupling is negligible, $\Delta_{SO}=0$, while the
Rashba parameter is nonzero, $\lambda_R\ne 0$. Equation~(\ref{6})
leads then to the following formula for the spin Hall
conductivity:
\begin{eqnarray}
\label{13} \sigma^{s_{z}}_{xy}=
-\,\int\frac{d\varepsilon}{2\pi}\frac{d^{2}\textbf{k}}{(2\pi)^{2}}
\frac{8 i e\, v^{2} \lambda^{2}_{R} \,(\varepsilon + \mu) [ v^{2}
(k^{2}_{x} - k^{2}_{y}) + (\varepsilon + \mu )^{2} ] }{\prod_{n =
1}^{4}[ \varepsilon  - E_{n}({\bf{k}}) + \mu + i \delta\, {\rm
sign}(\varepsilon ) ]}\nonumber\\\times\,\frac{1}{\prod_{m =
1}^{4}[ \varepsilon - E_{m}({\bf{k}}) + \mu +i \delta\, {\rm
sign}(\varepsilon )]},\hspace{0.5cm}
\end{eqnarray}
where $E_{i}({\bf{k}})$ ($i=1-4$) describe the electron energy
spectrum,
\begin{eqnarray}
E_{1}({\bf{k}}) = \lambda_{R} + (\lambda^{2}_{R} + v^{2}k^{2})^{1/2}, \\
E_{2}({\bf{k}}) = \lambda_{R} - (\lambda^{2}_{R} + v^{2}k^{2})^{1/2}, \\
E_{3}({\bf{k}}) = -\lambda_{R} + (\lambda^{2}_{R} + v^{2}k^{2})^{1/2}, \\
E_{4}({\bf{k}}) = -\lambda_{R} - (\lambda^{2}_{R} +
v^{2}k^{2})^{1/2}.
\end{eqnarray}
The states $E_{1}({\bf{k}})$ and $E_{3}({\bf{k}})$ correspond to
the conduction bands while $E_{2}({\bf{k}})$ and $E_{4}({\bf{k}})$
to the valence bands. We consider first the case when $|\mu| >
2\,\lambda_{R}$.

\subsection{The case of $|\mu| > 2\,\lambda_{R}$}

When the Fermi level is in the two valence bands, $\mu <
-2\,\lambda_{R}$, then upon integrating Eq.(\ref{13}) over
$\varepsilon$ one arrives at the following formula
\begin{equation} \label{14} \sigma^{s_{z}}_{xy}=\frac{e\,v^{2}}{16
\pi \lambda_{R}}\int d k\, \frac{2 \lambda^{2}_{R} k + v^{2} k^{3}
}{(v^{2}k^{2} + \lambda^{2}_{R})^{3/2}}\,
[f(\varepsilon_{2})-f(\varepsilon_{4})],
\end{equation}
where $f(\varepsilon)$ is the Fermi-Dirac distribution (assumed
here for $T=0$). Taking now into account the notation introduced
in Eq.(10), one finds
\begin{equation}
\sigma^{s_{z},0}_{xy} = 0.
\end{equation}
and
\begin{equation}
\delta \sigma^{s_{z}}_{xy} =  - \frac{e}{4 \pi}
\frac{\mu^{2}}{2(\mu^{2} - \lambda_{R}^{2})}\, .
\end{equation}
As before, $\sigma^{s_{z},0}_{xy}$ is the contribution from both
fully occupied valence bands, while $\delta \sigma^{s_{z}}_{xy}$
takes into account the empty part of the valence bands. Now, the
contributions from the two fully occupied valence bands cancel
each other, so $\sigma^{s_{z},0}_{xy}$ vanishes exactly.

When the Fermi level is in both conduction bands, $\mu
> 2\,\lambda_{R}$, the spin Hall conductivity can be calculated in a similar way,
and the formula for $\delta \sigma^{s_{z}}_{xy}$ takes the same
form as for $\mu < 2\,\lambda_{R}$, i.e. Eq.(22). Thus, the total
spin Hall conductivity for $|\mu| > 2\,\lambda_{R}$ can be written
as
\begin{equation}
\sigma^{s_{z}}_{xy} =  \frac{\mu^{2}}{2(\mu^{2} -
\lambda_{R}^{2})}\,\frac{e}{4 \pi}
\end{equation}
for the Fermi level in conduction bands, $\mu > 2\,\lambda_{R}$,
and
\begin{equation}
\sigma^{s_{z}}_{xy} =  - \frac{\mu^{2}}{2(\mu^{2} -
\lambda_{R}^{2})}\,\frac{e}{4 \pi}
\end{equation}
for the Fermi level in the valence bands, $\mu < 2\,\lambda_{R}$.

\begin{figure}
\includegraphics[width=0.47\textwidth,angle=0]{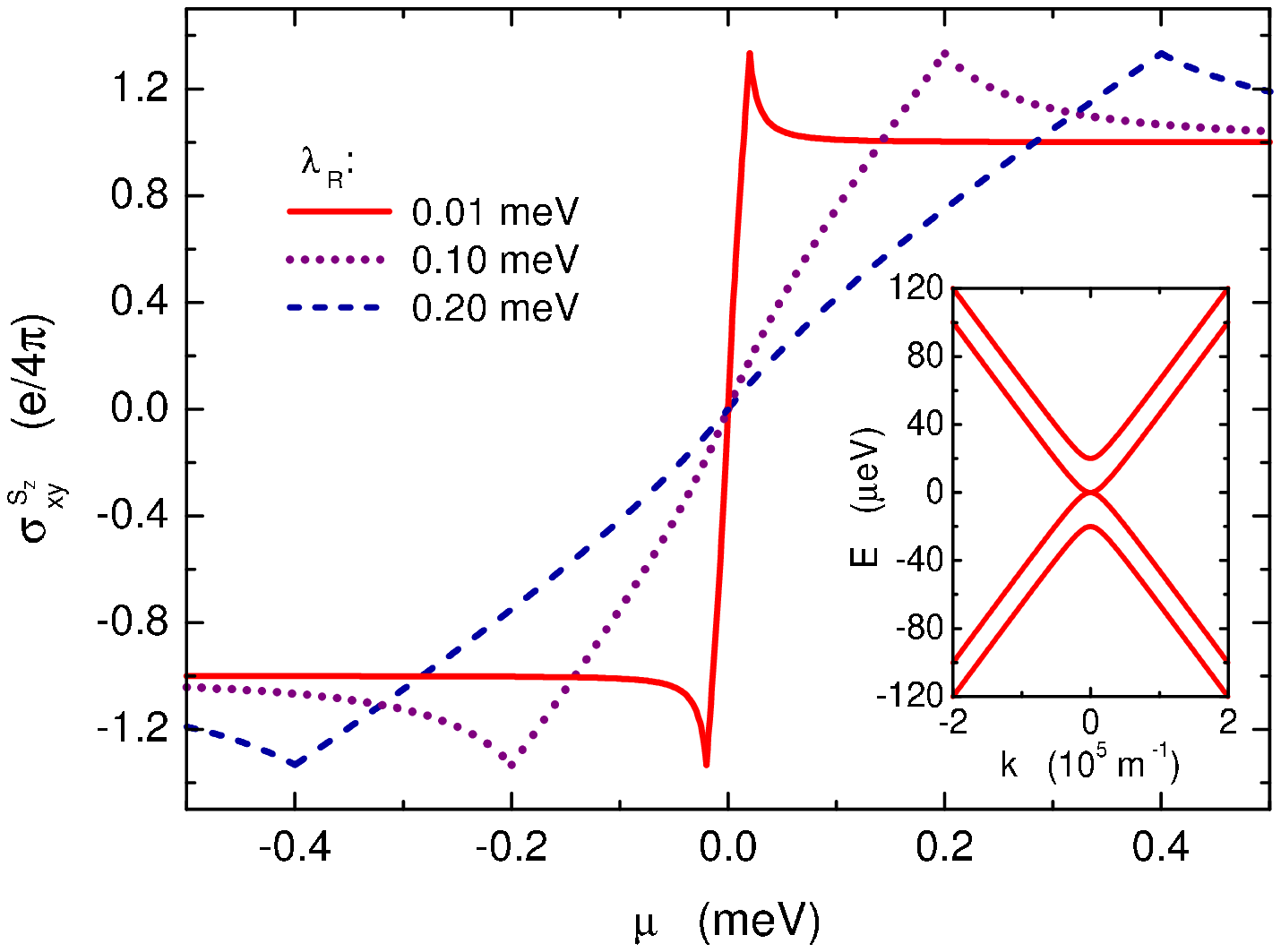}
\caption{(color online) Spin Hall conductivity in the absence of
intrinsic spin-orbit interaction and for indicated values of the
Rashba coupling parameter $\lambda_{R}$. The other parameters are
as in Fig.1. The inset shows the energy spectrum in the vicinity
of the energy gap for $\lambda_{R}=0.01$meV. }
\end{figure}

\subsection{The case of $|\mu| < 2\,\lambda_{R}$}

Now, we assume that $|\mu| < 2\,\lambda_{R}$. The only difference
is that now the Fermi level is either in one valence band or in
one conduction band. Spin Hall conductivity can be calculated in a
similar way as before and one finds
\begin{equation}
\label{20} \sigma^{s_{z}}_{xy} = \frac{\mu (\mu +
2\,\lambda_{R})}{4\, \lambda_{R}(\mu + \lambda_{R})}\,\frac{e}{4
\pi}
\end{equation}
for the Fermi level in the conduction band, and
\begin{equation}
\label{21} \sigma^{s_{z}}_{xy} = \frac{\mu (\mu -
2\,\lambda_{R})}{4\, \lambda_{R}(\mu - \lambda_{R})}\,\frac{e}{4
\pi}
\end{equation}
for the Fermi level in the valence band.

Figure 2 presents numerical results for the spin Hall conductivity
as a function of the chemical potential. As before, we included a
factor of 2 in order to take into account contribution from the
second Dirac point. We note, that now the spin Hall conductivity
tends to $\sigma^{s_{z}}_{xy}=-e/4\pi$ in the limit of $\mu\to
-\infty$, while for $\mu\to \infty$ it tends to the
$\sigma^{s_{z}}_{xy}=e/4\pi$. This behavior is different from that
for intrinsic spin-orbit interaction, where the contribution from
conduction bands cancelled the corresponding part from the valence
bands. It is also worth to note that now  the spin Hall
conductivity is antisymmetric with respect to change of the Fermi
level sign.

We have also checked the other components of the spin Hall
conductivity. As before, these components vanish exactly,
$\sigma^{s_{x}}_{xy} = \sigma^{s_{y}}_{xy} =0$.

\section{The case with $\lambda_{R} \neq 0$ and $\Delta_{SO} \neq 0$}

When both intrinsic and Rashba spin-orbit interactions are present
in the system, their interplay leads to interesting and peculiar
behavior of the spin Hall conductivity. The analytical formulas,
however, are much to complex to be presented here, so we will show
mainly results of numerical calculations.

Writing the Green function as
\begin{equation}
G_{\mathbf
k}(\varepsilon)=\frac{g_{\bf{k}}(\varepsilon)}{{\prod_{n = 1}^{4}[
\varepsilon  - E_{n}({\bf{k}}) + \mu + i \delta\, {\rm
sign}(\varepsilon ) ]}}\, ,
\end{equation}
where $E_n({\mathbf k})$ ($n=1-4$) are the dispersion relations
for the conduction and valence bands,
\begin{eqnarray}
E_{1}({\bf{k}}) = \lambda_R + \sqrt{(\Delta_{SO} - \lambda_R)^{2} + v^{2}k^{2}},\hspace{0.3cm} \\
E_{2}({\bf{k}}) = \lambda_R - \sqrt{(\Delta_{SO} - \lambda_R)^{2} + v^{2}k^{2}},\hspace{0.3cm} \\
E_{3}({\bf{k}}) = -\lambda_R + \sqrt{(\Delta_{SO} + \lambda_R)^{2} + v^{2}k^{2}}, \\
E_{4}({\bf{k}}) = -\lambda_R - \sqrt{(\Delta_{SO} + \lambda_R)^{2}
+ v^{2}k^{2}},
\end{eqnarray}
and taking into account the explicit form of $v_x$ and $v_y$ one
may write the relevant trace in Eq.(8) as
\begin{eqnarray} \label{22} Tr \left\{
\left(\begin{array}{cc}
                     0 & \sigma_{z} \\
                     \sigma_{z} & 0 \\
                   \end{array}
                 \right)g_{\textbf{k}}(\varepsilon_{1})\left(
              \begin{array}{cc}
                0 & -I \\
                I & 0 \\
              \end{array}
            \right)g_{\textbf{k}}(\varepsilon_{2})\right\} =
            \mathcal{P}(\varepsilon_{1},\,\varepsilon_{2}),
\end{eqnarray}
where $\mathcal{P}(\varepsilon_{1},\,\varepsilon_{2})$ is a
certain function of $\varepsilon _1=\varepsilon +\mu +\omega $
and $\varepsilon _2=\varepsilon +\mu $. Expanding
$\mathcal{P}(\varepsilon_{1},\,\varepsilon_{2})$ as
\begin{eqnarray}
\label{72} \mathcal{P}(\varepsilon_{1},\,\varepsilon_{2})=\left.
\mathcal{P}
(\varepsilon_{1},\,\varepsilon_{2})\right|_{\varepsilon_{1} =
\varepsilon_{2} = \varepsilon + \mu}\hskip3cm\nonumber\\ +\omega\,
\left.\frac{\partial \, \mathcal{P}
(\varepsilon_{1},\,\varepsilon_{2})}{\partial
\varepsilon_{1}}\right|_{\varepsilon_{1} = \varepsilon_{2} =
\varepsilon + \mu} + ... \,.
\end{eqnarray}
and taking into account that
$\left.\mathcal{P}(\varepsilon_{1},\,\varepsilon_{2})\right|_{\varepsilon_{1}
= \varepsilon_{2} = \varepsilon + \mu} = 0$, one finds
\begin{eqnarray}
\label{73} \mathcal{P}(\varepsilon+\omega ,\,\varepsilon) \cong
-\omega\left\{-4[(k^{2}v^{2} + \Delta_{SO}^{2})^{2} +
4v^{2}\lambda_R^{2}(k^{2}_{x} - k^{2}_{y})]\right. \nonumber\\
+ 16 \lambda_R^{2} [\Delta_{SO}^{2} + v^{2} (k^{2}_{x} -
k^{2}_{y})](\varepsilon + \mu)
 \nonumber\\+ 8\Delta_{SO}(v^{2}k^{2} + \Delta_{SO}^{2} - 4\lambda_R^{2})(\varepsilon + \mu)^{2}
 \nonumber\\\left. + 16\lambda_R^{2}(\varepsilon + \mu)^{3} - 4\Delta_{SO} (\varepsilon + \mu)^{4} \right\}.
 \hskip0.5cm
\end{eqnarray}

These formulas can be then used to calculate spin Hall
conductivity, either analytically by integrating over $\epsilon$
and $\mathbf k$ or numerically. Since analytical formula are
generally rather cumbersome, we performed numerical calculations,
while analytical formula will be presented only for some special
cases (see below).

\subsection{$\lambda_{R} > \Delta_{SO}$}

\begin{figure}[h]
\includegraphics[width=0.47\textwidth,angle=0]{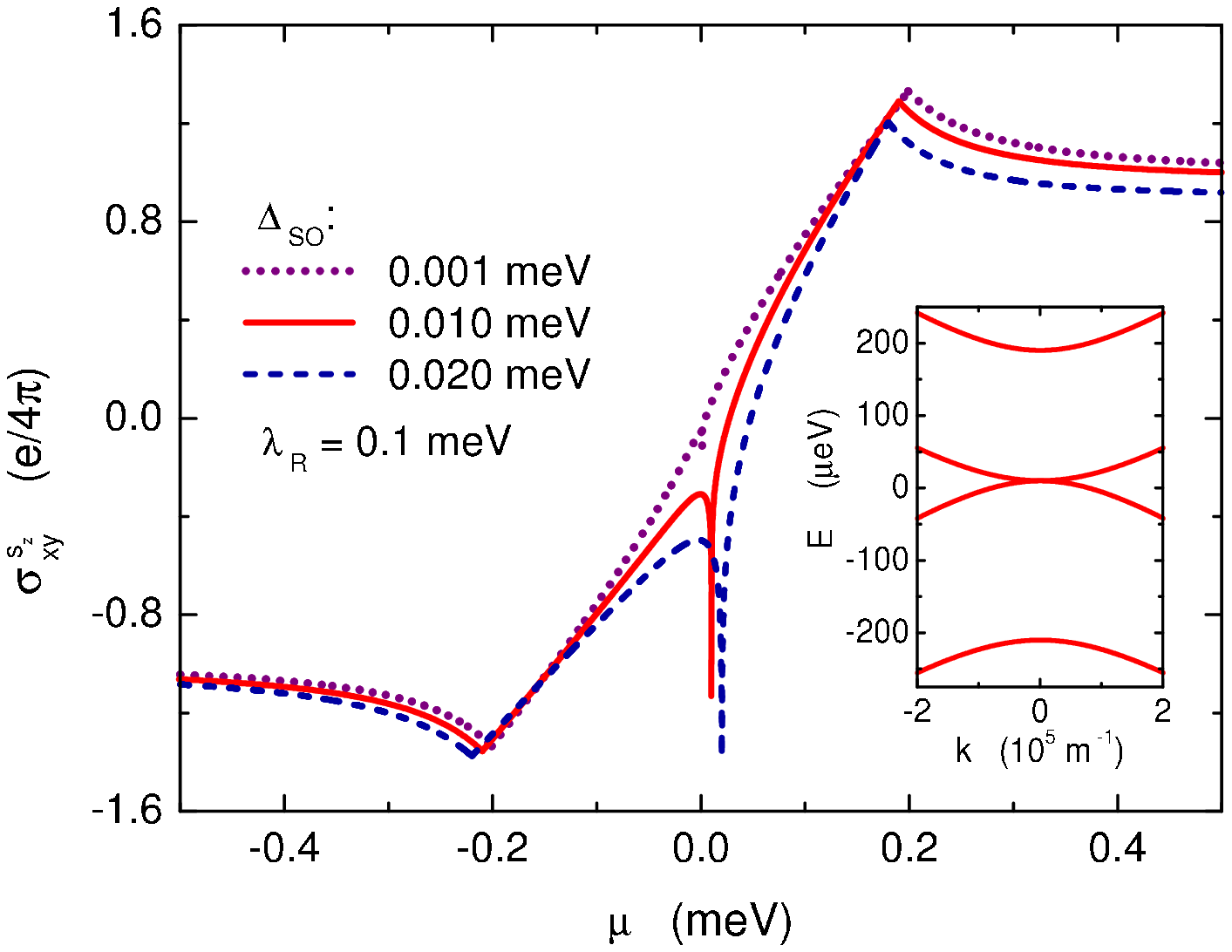}
\caption{(color online) Spin Hall conductivity for
$\lambda_R=0.1$meV and indicated values of $\Delta_{SO}$,
$\Delta_{SO}<\lambda_R$. The inset shows the energy spectrum
in the vicinity of the Dirac point
for $\Delta_{SO}=0.01$meV. The parameter $v$ is taken as in
Fig.1.}
\end{figure}

Consider first the case when $\lambda_R$ is significantly larger
than $\Delta_{SO}$. The corresponding spin Hall conductivity is
shown in Fig.~3 for $\lambda_R=0.1$ meV and three different values
of $\Delta_{SO}$ (smaller than $\lambda_R$). When $\Delta_{SO}$ is
much smaller than $\lambda_R$, then the conductivity is determined
practically only by the Rashba coupling. With increasing
$\Delta_{SO}$, the interplay of both interactions leads to
anomalous behavior of the spin Hall conductivity. More
specifically, the conductivity becomes diverging when the Fermi
level approaches the point, at which the top of the upper valence
band touches the bottom of the lower conduction band (see the
inset in Fig.3). This appears when $\mu =\Delta_{SO}$. Asymptotic
behavior of the conductance near the point $\mu = \Delta_{SO} $ is
described by the term $(\Delta_{SO}/2 \lambda_R ) \; \ln
[-\Delta_{SO} - \lambda_R + \sqrt{(\lambda_R + \mu)^{2}}]$ when
$\mu = \Delta_{SO} $ is approached from the right ($\mu
>\Delta_{SO}$) side, and $(\Delta_{SO}/2 \lambda_R ) \;
\ln[\Delta_{SO} - \lambda_R + \sqrt{(\lambda_R - \mu)^{2}}]$ when
it is  approached  from the left ($\mu <\Delta_{SO}$) side.

\subsection{$\lambda_{R} < \Delta_{SO}$}

Let us now consider the opposite situation, when $\lambda_R$ is
smaller than $\Delta_{SO}$. The corresponding spin Hall
conductivity is shown in Fig.4 for $\lambda_R=0.01$ meV and three
different values of $\Delta_{SO}$ (larger than $\lambda_R$).
General shape of the curve showing spin Hall conductivity as a
function of the chemical potential is similar to that for
$\lambda_R=0$. However, the interplay of intrinsic spin-orbit
interaction and Rashba coupling leads to an interesting feature.
More specifically, there is now no divergence, but a weak kink in
the conductance appears on the negative chemical potential side.
When $\Delta_{SO}$ decreases and approaches $\lambda_R$, the kink
becomes more pronounced. The kink is associated with splitting of
the valence band by the Rashba interaction. The upper valence band
edges are now at $-\Delta_{SO}-2\lambda_R$ and
$-\Delta_{SO}+2\lambda_R$.

\begin{figure}[h]
\includegraphics[width=0.47\textwidth,angle=0]{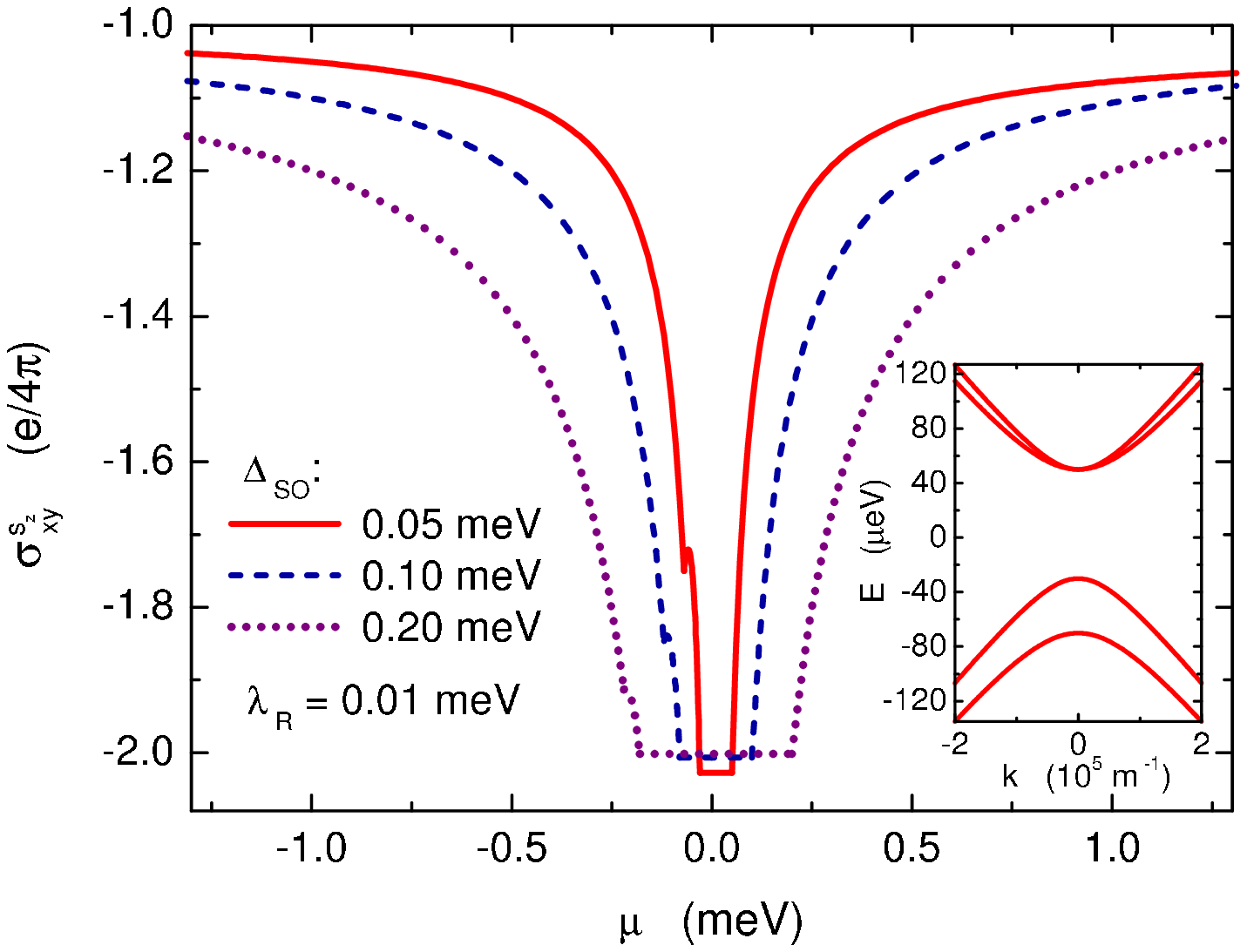}
\caption{(color online) Spin Hall conductivity for
$\lambda_R=0.01$meV and indicated values of $\Delta_{SO}$,
$\Delta_{SO}>\lambda_R$. The inset shows the energy spectrum in
the vicinity of the Dirac point for $\Delta_{SO}=0.05$meV. The
parameter $v$ is taken as in
Fig.1.}
\end{figure}

\subsection{$\lambda_{R} = \Delta_{SO}$}

Variation of the spin Hall conductivity with the Fermi level
becomes more complex when both $\Delta_{SO}$ and $\lambda_R$ are
comparable. Some simple analytical results, however, can be
obtained for $\lambda_R=\Delta_{SO}$. For this particular case,
the bottom edges of two conduction bands coincide with top edges
of one of the valence band, while the top edge of the second
valence band is much below (see the inset in Fig.5).

The relevant formula for the spin Hall conductivity depends then
on the Fermi level as follows:

\subsubsection{$\mu < -3\Delta_{SO}$}

For $\mu < -3\Delta_{SO}$, the spin Hall conductivity is given by
the formula
\begin{equation}
\sigma^{S_{z}}_{xy} = -\frac{\mu}{\mu + \Delta_{SO}}
\frac{e}{4\pi}
\end{equation}
This formula covers the energy range up to the top edge of the
lower valence band.

\subsubsection{$-3\Delta_{SO}< \mu < \Delta_{SO}$}

When $-3\Delta_{SO}< \mu < \Delta_{SO}$, the corresponding formula
for the conductivity takes the form
\begin{equation}
\sigma^{S_{z}}_{xy} = \left[ \frac{\mu}{2 \Delta_{SO}} +
\ln\left(\frac{\Delta_{SO} - \mu}
{4\Delta_{SO}}\right)\right]\frac{e}{4\pi}.
\end{equation}
This formula, in turn, describes spin Hall conductivity when the
chemical potential is between the top edge of the lower valence
band and bottom edges of the conduction bands. Note, that this
formula leads to diverging spin Hall conductivity when the Fermi
level tends from left (lower values) to $\Delta_{SO}$.

\subsubsection{ $\mu > \Delta_{SO}$}

Finally, for $\mu > \Delta_{SO}$ the spin Hall conductivity is
equal to
\begin{equation}
\sigma^{S_{z}}_{xy} = \frac{\mu}{\mu + \Delta_{SO}} \frac{e}{4\pi}
\end{equation} This
formula gives a finite spin Hall conductivity in the whole range
of its applicability, also at the point $\mu = \Delta_{SO}$.

\begin{figure}[h]
\includegraphics[width=0.47\textwidth,angle=0]{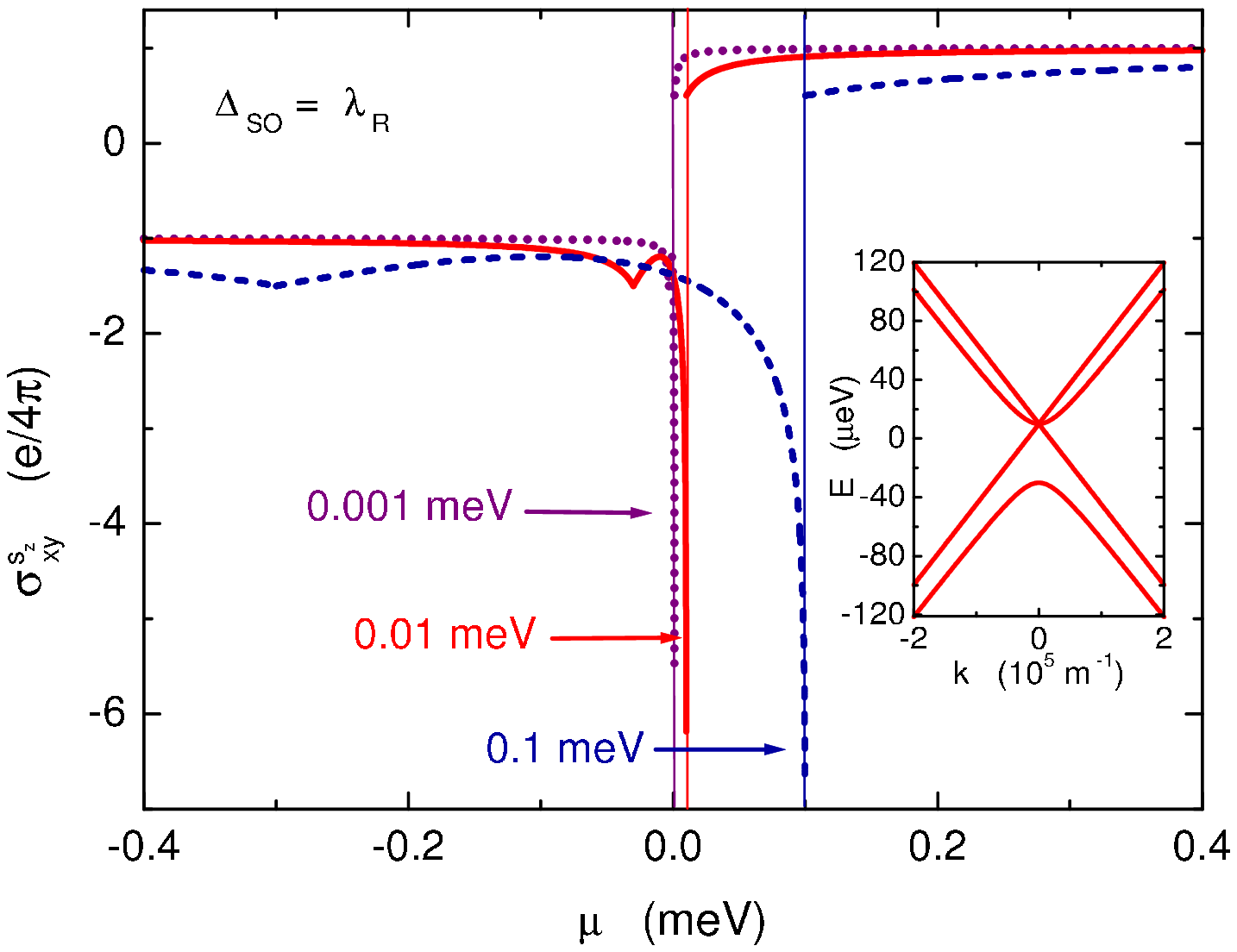}
\caption{(color online) Spin Hall conductivity for indicated
values of  $\lambda_R=\Delta_{SO}$. The inset shows the energy
spectrum in the vicinity of the Dirac point for
$\lambda_R=\Delta_{SO}=0.01$meV. The parameter $v$ is taken as in
Fig.1. The vertical dotted lines indicate the position where the
anomaly appears.}
\end{figure}

The spin Hall conductivity for $\lambda_R=\Delta_{SO}$ is shown in
Fig.5 for indicated values of $\lambda_R=\Delta_{SO}$. The anomaly
at $\mu = \lambda_R=\Delta_{SO}$ is now clearly visible. When the
chemical potential $\mu$ tends to $\mu = \lambda_R=\Delta_{SO}$
from the right ($\mu >\Delta_{SO}$) side, the conductivity is
finite, while when it tends to $\mu = \lambda_R=\Delta_{SO}$ from
the left ($\mu ;\Delta_{SO}$) side, the conductivity becomes
diverging. The vertical dotted lines in Fig.~5 indicate only the
position where the anomaly appears.

\section{Summary and discussion}

Assuming intrinsic and Rashba spin-orbit interaction we have
calculated topological contribution to the spin Hall conductivity.
In the limit of vanishing Rashba term we arrived at the results
which are in agreement with those available in the relevant
literature. When, in turn, the Rashba coupling dominates and
intrinsic spin orbit-coupling vanishes, we have found asymmetric
behavior of the spin Hall conductivity with respect to the sign
reversal of the chemical potential. Such a change in the chemical
potential can be achieved with a gate voltage, for instance.

When both intrinsic and Rashba spin-orbit interactions are
present, their interplay leads to some peculiar and anomalous
behavior of the spin hall conductivity with the Fermi level. In
particular, for some range of spin-orbit parameters, the spin Hall
conductivity was found to diverge when the Fermi level approaches
the limit $\mu = \Delta_{SO}$.

\section*{Acknowledgements}
This work was supported by the EU grant CARDEQ under contract
IST-021285-2, FCT Grant PTDC/FIS/70843/2006 in
Portugal, and by funds from the Ministry of Science and Higher
Education as a research project in years 2007 -- 2010.

\end{document}